\documentclass[twocolumn,superscriptaddress,showpacs,preprintnumbers,amsmath,amssymb,prb]{revtex4}
\usepackage{color}
\usepackage{graphicx}
\usepackage{dcolumn}
\usepackage{bm}
\newcommand{\xone} {{\mbox{Co${}_3$O${}_4$}}}

\newcommand{\xnul} {{\mbox{CoAl${}_2$O${}_4$}}}
\newcommand{\CoAlO} {{\mbox{Co(Al${}_{1-x}$Co${}_{x}$)${}_2$O${}_4$}}}

\newcommand{\twothird}{{\ensuremath{\frac{2}{3}}}}

\begin{document}
\preprint{APS}

\title{Evolution of magnetic states in frustrated diamond lattice antiferromagnetic ${\CoAlO}$ spinels}
\author{O. Zaharko,\cite{now}}
\affiliation{Laboratory for Neutron Scattering, ETHZ \& PSI, CH-5232 Villigen, Switzerland}
\author{A. Cervellino,}
\affiliation{Laboratory for Neutron Scattering, ETHZ \& PSI, CH-5232 Villigen, Switzerland}
\affiliation{Swiss Light Source, Paul Scherrer Institute, CH-5232 Villigen, Switzerland}
\author{V. Tsurkan,}
\affiliation{Experimental Physics V, Center for Electronics Correlations and Magnetism, University of Augsburg, D-86159 Augsburg, Germany}
\affiliation{Institute of Applied Physics, Academy of Sciences of Moldova, MD-2028 Chisinau, Republic of Moldova}
\author{N. B. Christensen,}
\affiliation{Laboratory for Neutron Scattering, ETHZ \& PSI, CH-5232 Villigen, Switzerland}
\affiliation{Materials Research Division, Ris\o~Nat. Lab. for Sustainable Energy, Technical University of Denmark}
\affiliation{Nano-Science Center, Niels Bohr Institute, University of Copenhagen, DK-2100 Copenhagen, Denmark}
\author{and A. Loidl}
\affiliation{Experimental Physics V, Center for Electronics Correlations and Magnetism, University of Augsburg, D-86159 Augsburg, Germany}
\date{\today}

\begin{abstract}
Using neutron powder diffraction and Monte-Carlo simulations we show that a spin-liquid regime emerges at $\it{all}$ compositions in the diamond-lattice antiferromagnets ${\CoAlO}$. This spin-liquid regime induced by frustration due to the second-neighbour exchange coupling $J_{2}$,
is gradually superseded by antiferromagnetic collinear long-range order ($\bf{k}$=0) at low temperatures. Upon substitution of Al$^{3+}$ by Co$^{3+}$ in the octahedral B-site the temperature range occupied by the spin-liquid regime narrows and T$_N$ increases. To explain the experimental observations we considered magnetic anisotropy $D$ or third-neighbour exchange coupling $J_{3}$ as degeneracy-breaking perturbations. We conclude that ${\CoAlO}$ is below the theoretical critical point $J_{2}/J_{1}$=1/8, and that magnetic anisotropy assists in selecting a collinear long-range ordered ground state, which becomes more stable with increasing  x due to a higher efficiency of O-Co$^{3+}$-O  as an interaction path compared to O-Al$^{3+}$-O.\\
\end{abstract}

\pacs{75.50.Mm, 61.05.F-}
\keywords{spin liquid, neutron scattering, spinels}
\maketitle

\section{Introduction}

Magnetic systems with frustration induced by competing exchange interactions quite often manifest unconventional ground states, the most intriguing of which are spin-liquids\cite{note1}. Recently one such exotic state, a 'spiral spin-liquid', was uncovered theoretically in a classical treatment of diamond-lattice Heisenberg antiferromagnets (AFM) by Bergman {\it et al.}\cite{Bergman07} who showed that competition between nearest and next-nearest neighbor 
exchange couplings $J_{1}$ and $J_{2}$  creates - for $J_{2}/J_{1}>$1/8 - a highly degenerate ground state consisting of a set of coplanar spirals, whose propagation vectors form a continuous surface in momentum space. The frustration results in a rich phase diagram as a function of the ratio $J_{2}/J_{1}$. The degeneracy of these ground states can be lifted by thermal\cite{Bergman07} or quantum\cite{Bernier08} fluctuations leading to an 'order-by-disorder' phase transition from a spiral spin-liquid regime to an ordered state.\\

Among the diamond-lattice AFM, compounds with the spinel structure recently attracted much attention\cite{Fritsch04, Krimmel06a, Krimmel09}. In particular, Co-Al oxides were considered as promising candidates for study of 'order-by-disorder' physics \cite{Bergman07, Bernier08}. 
In these compounds of general stoichiometry AB$_2$O$_4$ the tetrahedral A-sites are occupied by high-spin (S=3/2) magnetic Co$^{2+}$ ions which form a diamond lattice consisting of two interpenetrating face-centered cubic sublattices coupled antiferromagnetically. The octahedral B-sites can be filled either by nonmagnetic Al$^{3+}$ ions and/or by low-spin (S=0) nonmagnetic Co$^{3+}$ ions.\\

Existing experiments on Co-Al oxide spinels do not provide a clear picture. An early neutron-diffraction study\cite{Roth63} on ${\xone}$ showed that the magnetic moments of the Co$^{2+}$ ions located at the tetrahedral sites form a simple collinear AFM below the Neel temperature T$_N$=40 K. This picture has been questioned by a recent $\mu$SR study\cite{Ikedo07} which found two frequency components near T$_N$ suggesting incommensurate magnetic order. Experimental observations on ${\xnul}$ are also contradictory. A powder neutron diffraction study of Roth\cite{Roth64} suggested long-range AFM order below 4 K, while Krimmel {\it et al.}\cite{Krimmel06} detected a spin liquid (or glassy-like) ground state. Electron spin resonance (ESR), magnetization and specific heat measurements identified the ground state of ${\xnul}$ as spin-glass-like with a high frustration parameter $\mid$T$_{CW}$$\mid$/T$_N$ of 22 [Ref.\onlinecite{Tristan05}] and 10 [Ref.\onlinecite{Suzuki07}], respectively. These experimental results are consistent with the calculations of Bergman {\it et al.}\cite{Bergman07} which placed ${\xnul}$ in the region of $J_{2}/J_{1} \approx$ 1/8 where the spiral surface begins to develop. To clarify the situation, it is essential to understand whether the ground state of ${\xone}$ is a simple collinear antiferromagnet and if ${\xnul}$ is a spiral spin-liquid fluctuating among degenerate spirals. It is also important to understand why substitution within the nonmagnetic B-site changes the magnetic properties so drastically. Tristan {\it et al.}\cite{Tristan08} attempted to answer these questions based on bulk macroscopic measurements of ${\CoAlO}$ system. They proposed that the spin-liquid state is realized in ${\xnul}$ (x=0), while with increasing Co substitution x the second neighbor coupling $J_{2}$ decreases and collinear AFM long-range order develops.\\
In this work we address these important issues and refine the evolution of the magnetic states in ${\CoAlO}$ spinels by means of neutron powder diffraction supported by Monte-Carlo simulations\cite{Metropolis49}. Our analysis reveals that for x=0 the system is close to the critical point $J_{2}/J_{1}$=1/8 in the phase diagram while frustration is weaker for x$>$0. We argue that magnetic anisotropy assists in the selection of a collinear long-range ordered ground state. The difference in T$_N$ and in the extent of the spin-liquid regime apparently originates from the decrease of the $J_{2}/J_{1}$ ratio with increasing x.\\
Polycrystalline ${\CoAlO}$ samples with x=0, 0.35, 0.75 and 1 have been prepared as reported in Ref.~\onlinecite{Tristan05}. 
Thorough structural characterizations by high-resolution x-ray synchrotron diffraction (on the Materials Science beamline at Swiss Light Source, $\lambda$=0.41414~\AA) and neutron diffraction (on the HRPT diffractometer at the Swiss Neutron Spallation Source SINQ, $\lambda$=1.1545~\AA) indicated 
the absence of inversion in all samples excepting ${\xnul}$. For the latter compound the inversion is 17\% and this might influence T$_N$.
Analyzing the lattice constants and x-ray peak profiles we concluded that
the Co$^{3+}$ ions are homogeneously and randomly distributed over the B-sites. This follows because 
(i) the lattice constants obey Vegard's law as the function of x and (ii) there is no peak asymmetry, which would have
been present in the high-resolution data if any significant degree of inhomogeneity was present.\\
\begin{figure}[tbh]
\includegraphics[width=86mm,keepaspectratio=true]{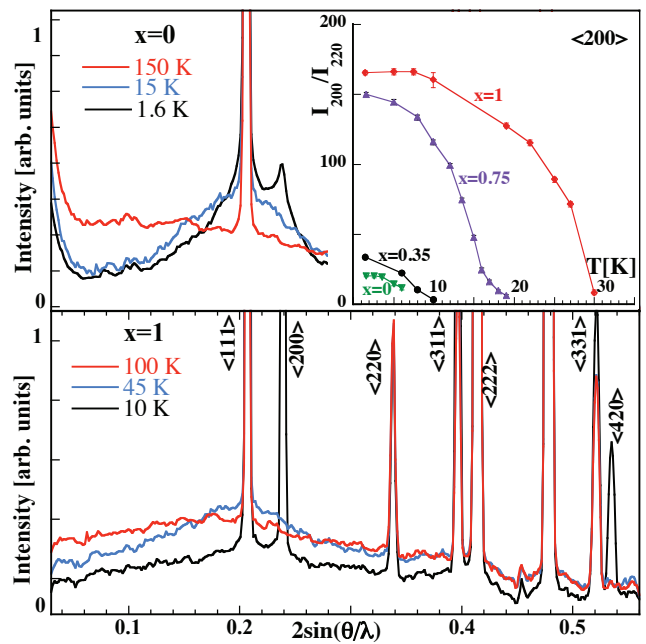}
\caption {(Color online) DMC patterns of x=0 and x=1 samples at selected temperatures. Reasonable statistics on diffuse scattering has been obtained by measuring for about 10 h at one temperature. Inset: Temperature dependence of intensity of the $<$200$>$ magnetic reflection normalized to the nuclear $<$220$>$ intensity.}
\label{fig1}
\end{figure}
Medium-resolution neutron powder diffraction patterns for four compositions have been collected in the 
temperature range 1.5 K - 150 K on the DMC instrument at SINQ with a neutron wavelength $\lambda$=2.4526~\AA. 
For all compositions, broad bumps resembling a liquid-like structure factor and reflecting short-range magnetic order start to develop below the Curie-Weiss temperature $|\theta_{CW}|$=110 K\cite{Tristan08};
they narrow and shift to higher sin$\theta/\lambda$ with cooling (Fig.~\ref{fig1}). 
Approaching T$_N$ the diffuse scattering localizes near the $<$111$>$ and $<$200$>$ positions. Below T$_N$ the 
liquid-like features remain but gradually loose spectral weight as magnetic Bragg peaks due to long-range order develop.  
The spectral weight of the  diffuse scattering component continuously increases from ${\xone}$ to ${\xnul}$ (Fig.~\ref{fig2}) as the frustration parameter $\mid$T$_{CW}$$\mid$/T$_N$ grows. We can quantify it by the area of a Lorentzian fitted to the first diffuse bump. As shown in the inset of Fig.~\ref{fig2} this area is largest roughly at  T/T$_N \approx$1 for all compositions, but for x=0 and x=0.35 diffuse scattering develops far above and remains significantly below this value. 
For x=1 the T/T$_N$ interval revealing diffuse scattering is narrower but still much to large to be interpreted as classical critical 
scattering of a 3D long-range ordered magnet.
 
Regarding the long-range order, all samples show magnetic Bragg peaks at low temperatures. The ordering temperature T$_N$ and the static ordered magnetic moment decrease from ${\xone}$ to ${\xnul}$ (see Table ~\ref{tab1} and inset of Fig.~\ref{fig1}). It should be noted that the x=0 diffraction pattern does not correspond to a conventional long-range ordered state: diffuse scattering clearly dominates and the $<$200$>$ peak is so broad and weak that the ordered moment cannot be determined with high accuracy. Nevertheless, the magnetic Bragg pattern is the same for all compositions and is consistent with the collinear two-sublattice model proposed by Roth\cite{Roth63}.\\
\begin{table}
\caption{The N{\'e}el temperature, T$_N$, and the ordered magnetic moment, M, determined at 1.6 K from the DMC patterns.}
\label{tab1}
\begin{ruledtabular}
\begin{tabular}{cccc}
x&T$_N$ [K]&M [$\mu_B$]\\
1.0&29&3.53(3)\\
0.75&16.5&2.59(3)\\
0.35&9&1.31(4)\\
0.0&5&0.25(7)\\
\end{tabular}
\end{ruledtabular}
\end{table}
To model the observed magnetic diffuse scattering we used the quasistatic approximation\cite{vanHove}, which assumes that the observed diffuse scattering can be attributed to static correlations. This approach is justified since the bandwidth of magnetic excitations is below 6 meV whereas all scattering with energies below 14.7 meV is summed in our diffraction experiment. 
\begin{figure}[tbh]
\includegraphics[width=86mm,keepaspectratio=true]{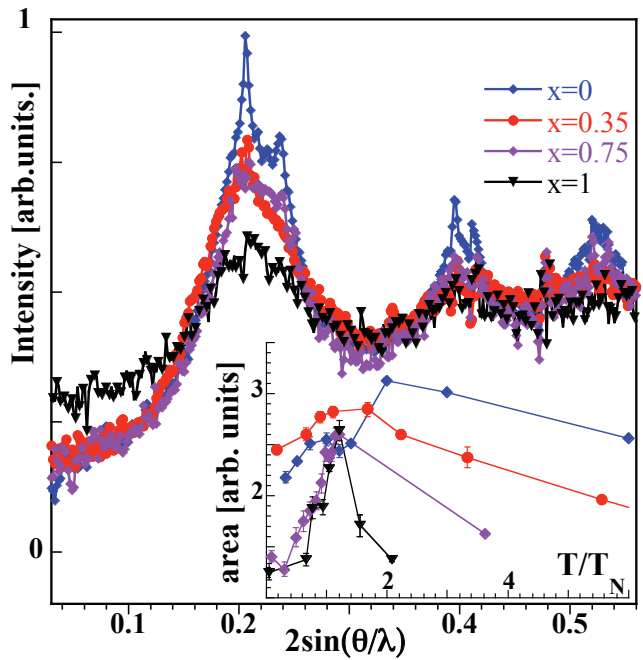}
\caption {(Color online) Maximal diffuse magnetic scattering in the differential T-150 K neutron powder patterns. The T-value is equal to 6, 15, 19, 45 K for the compositions x=0, 0.35, 0.75 and 1, respectively. Inset: Evolution of diffuse scattering quantified as an area of Lorentzian fitted to the first bump.}
\label{fig2}
\end{figure}
We performed a Monte-Carlo search for the ground state of a cluster of 1047 atoms. 
The energy of the classical Heisenberg Hamiltonian for spins ${\bf{S}}$ interacting with first- and second-neighbour antiferromagnetic couplings $J_{1},J_{2} >0$ 
\begin{equation}
H = J_{1} \mathop{\sum}_{<ij>} {\bf{S}}_{i} \cdot {\bf{S}}_{j}+ J_{2} \mathop{\sum}_{<<ij>>} {\bf{S}}_{i} \cdot {\bf{S}}_{j}
\label{eq1}
\end{equation}
was minimized 
($<ij>$ means that the sum runs over first-neighbour pairs, $<<ij>>$ on second-neighbours pairs). 
If we add either a third-neighbour interaction or an anisotropy contribution, the Hamiltonian will be completed with an additional
term $\Delta H$, given, respectively, as
\begin{equation}
\Delta H = J_{3} \mathop{\sum}_{<<<ij>>>} {\bf{S}}_{i} \cdot {\bf{S}}_{j} \qquad\text{(third-neighbour)}; 
\label{eq1-b}
\end{equation}
\begin{equation}
\Delta H =D\mathop{\sum}_{i}( {\bf{S}}_{i} \cdot{\bf{u}})^2 \qquad\text{(anisotropy)}.
\label{eq1-c}
\end{equation}
where $<<<ij>>>$ refers to third-neighbours pairs and ${\bf{u}}$ is the
anisotropy direction, that we took as (111). $D$ is the magnitude of the anisotropy term.
The moments were kept equal and constant in magnitude; their direction was 
changed at random, one at the time, and to obtain the ground state 
only energy-decreasing moves were accepted in the final stage. The stopping criterion for the 
Monte-Carlo was that the last 1000 accepted configurations had an energy spread less than 10$^{-6}$ $J_{1}$. 
Runs have been repeated for the whole range 0$\le J_{2}/J_{1}\le$1.
We calculated the static spin-pair correlation functions and diffraction patterns for the Monte-Carlo ground state (MCGS) and the reference AFM clusters for each $J_{2}/J_{1}$.
The correlation function is given by
\begin{equation}
CF(d) = {\twothird} \mathop{\sum}_{ij}{\bf{S}}_{i} \cdot {\bf{S}}_{j} \delta(|r_i-r_j|-d)
\label{eq2}
\end{equation}
where $d$ is the distance between spins at positions $r_i$ and $r_j$.
We found that for the range 0$<J_{2}/J_{1}<$1/8 the collinear AFM is the ground state, in agreement with Ref.~\onlinecite{Bergman07}. 
Remarkably, the MCGS was always configurationally different from a collinear AFM state even
if the energy difference was negligible (order of 1 mK or less). Analytical calculations (Fig.~\ref{fig3} inset), similar to Ref.~\onlinecite{Bergman07}, support this result. 
In fact, the energy minimum corresponding to the ground state for $J_{2}/J_{1}<$1/8 is extremely flat around the $q$=0 point of the first Brillouin zone, and therefore, 
very many states are in the range of thermal excitation at any accessible temperature.\\
The MCGS starts closely related to the collinear AFM state, but progressively departs from it with increasing $J_{2}/J_{1}$.
This is clearly seen in the decay of spin correlations, which can be quantified by the ratio $CF_{MCGS} / CF_{AF}$. 
In Fig.~\ref{fig3} we plot log$|CF_{MCGS}/CF_{AF}|$ versus squared distance $d^2$ between spins. 
Near $J_2/J_1$=1/8 the decay changes from predominantly Gaussian $e^{-1/2 \cdot(d/w)^2}$ (a straight line in this choice of axes) to a mixed Gaussian-exponential form $e^{-d/L} \cdot e^{-1/2 \cdot (d/w)^2}$ (a curve bending downwards for 200$< d^2 <$400 \AA$^2$). This behaviour can be attributed to the incipient spiral surface.
In the calculated diffraction patterns (Fig.~\ref{fig4}, top) these changes are evidenced as progressive broadening and distortion of Bragg peaks. 
Pure Gaussian correlation decay yields a Gaussian peak broadening, whereas gaussian-exponential decay yields Voigtian point spread (that is, with a sizable
Lorentzian convolution component).
For $J_2/J_1>$0.5 the correlations are completely changed and features of the collinear AFM state fade out.\\
\begin{figure}[tbh]
\includegraphics[width=86mm,keepaspectratio=true]{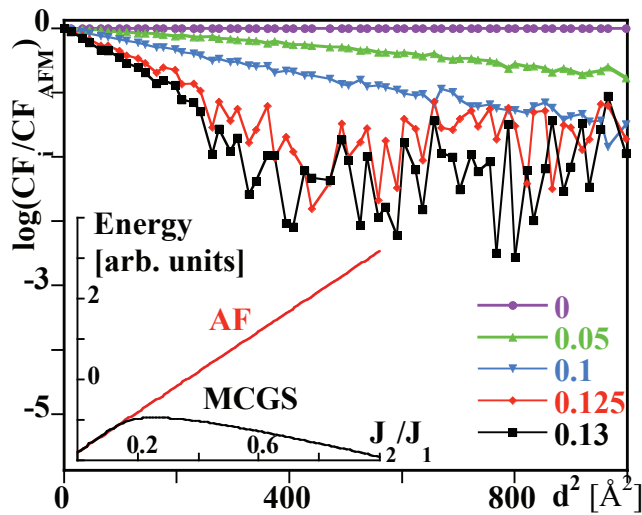}
\caption {(Color online) Log of ratio of spin-pair correlation functions $|CF_i/CF_{AFM}|$ versus $d^2$ for selected $J_2/J_1$ ratios.
Inset: The minimal energy curve for the collinear antiferromagnet (red) and for the best Monte-Carlo ground state (black).}
\label{fig3}
\end{figure}
We now compare the experimental and calculated diffuse scattering. The patterns shown in Fig.~\ref{fig2} were 
obtained as differences between data measured at  T= 6, 15, 19 and 45 K (for x=0, 0.35, 0.75 and 1, respectively) and 150 K.
In the calculated patterns in order to obtain the ordered AFM state at $T<T_N$ we needed to perturb the simple Heisenberg Hamiltonian so that the AFM minimum would be deeper and the AFM configration more stable. For this we considered separately two possible degeneracy-breaking perturbations: magnetic anisotropy and finite $J_3$ coupling. To retain a simple model we described anisotropy as a single-ion parameter $D$. The MC procedure was now changed, allowing for system equilibrations at different temperatures. 
All moves that would decrease the energy or that would increase it with probability $\propto e^{-\Delta E/T}$ were accepted\cite{note3}. 
When the average and fluctuations of the energy of a large number of the last accepted states were sufficiently stable the temperature was changed. 
Fixing $J_1$=1 as a convenient energy scale, $J_2$, $J_3$, $D$ or T were varied, each in several steps.\\
\begin{figure}[tbh]
\includegraphics[width=86mm,keepaspectratio=true]{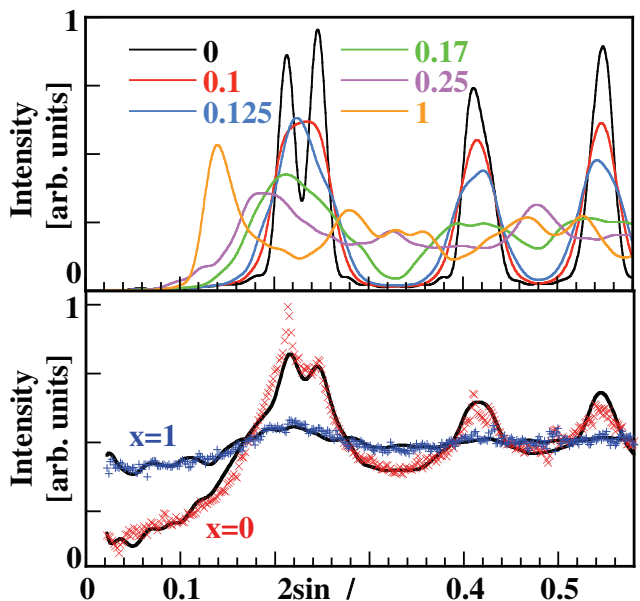}
\caption{(Color online) Top: Calculated powder diffraction patterns for several selected $J_2/J_1$ ratios at T = 0 K. 
Bottom: Observed (symbols) and calculated (line) powder diffraction patterns for x=0 and x=1 corresponding to the best model parameters. The calculated patterns are corrected for the magnetic form factor of Co$^{3+}$ and polynomial background.}
\label{fig4}
\end{figure}
Equally good fits were obtained with $J_3$ and magnetic anisotropy $D$ as perturbations, signifying that the available diffraction data are not sufficient to determine which of these terms stabilizes the ordered ground state. However, we give preference to the magnetic anisotropy following the theoretical study of  Ref.~\onlinecite{Lee08}. The best fits presented in Fig.~\ref{fig4} (bottom) suggest the ratio $J_{2}/J_{1}$=0.125 for $\xnul$ and $J_{2}/J_{1}$=0.05 for $\xone$. 
The change of the exchange energy with substitution apparently originates from the peculiarities of the electronic structure. Band structure analysis\cite{Walsh} shows that near the Fermi level in $\xone$ there are Co$^{3+}$ $d$ and oxygen $p$ states, while in $\xnul$ the Al $p$ states are absent from the Fermi level and the weight of O $p$ is diminished. This implies that the interaction path O-Co$^{3+}$-O is more effective and the corresponding exchange integrals are larger in $\xone$.\\ 
Our findings give a natural explanation of the $\mu$SR results\cite{Ikedo07} on $\xone$ as due to spin-liquid physics rather than incommensurate magnetic order. Also, based on the position of $\xnul$ in the
$J_{2}/J_{1}$ phase diagram, we suggest that even in an ideal  sample, with no inversion or other perturbation, the ground state would be a collinear AFM, though one might need very low temperatures to reach it.
As the ground state is the collinear AFM the general term 'spin-liquid' and not 'chiral spin-liquid' is appropriate for the high-temperature regime above T$_N$ in the title system.\\
A single crystal inelastic neutron scattering experiment would allow to extract the absolute values of $J_{1}, J_{2}$ and to validate our conclusions.
We remark that the elaborated approach to fit measured diffuse magnetic neutron scattering to Monte-Carlo simulations can be easily adapted to other frustrated systems and would be useful in justification of an anticipated Hamiltonian.\\
In summary we studied the evolution of magnetic states in ${\CoAlO}$ polycrystalline samples with temperature and substitution in the B-site. We observed short-range and long-range order for all compositions. Employing Monte-Carlo simulations we found that for x=0 the system is in the vicinity of the critical point $J_{2}/J_{1}$=1/8, where the spiral spin-liquid develops\cite{Bergman07}, but stays in the weakly frustrated limit for x$>$0. We also found that replacement in the nonmagnetic B-site changes the strength of exchange interactions which in turn leads to significant differences  in the ordering temperatures and in the extent of the spin-liquid regime.\\

We thank the expert experimental assistance of L. Keller and D. Sheptyakov.
The work was performed at SINQ and SLS, Paul Scherrer Insitute, Villigen, Switzerland. The support of PSI via the collaborative grant MOP1-33010-CH-08 of the GAP of the US CRDF is gratefully acknowledged.
NBC acknowledges the support by the Danish Natural Science Research Council under DANSCATT.


\begin{thebibliography}{99}
\bibitem{note1} Spin-liquid is characterized by rapid (at least exponential) decay of spin-spin correlations and the correlation length does not exceed the interatomic distance\cite{Canals98}, which result in a spin liquid-like structure factor.
\end{thebibliography}
\end{document}